\documentclass[utf8]{FrontiersinHarvard} % for articles in journals using the Harvard Referencing Style (Author-Date), for Frontiers Reference Styles by Journal: https://zendesk.frontiersin.org/hc/en-us/articles/360017860337-Frontiers-Reference-Styles-by-Journal
\usepackage{url,hyperref,lineno,microtype,subcaption}
\usepackage[onehalfspacing]{setspace}
\usepackage{graphicx}        % For eps figures, newer & more powerfull
\usepackage{color}           % For color text: \color command
\usepackage{breakurl}        % For breaking URLs easily trough lines
\usepackage[normalem]{ulem}

\chardef\us=`\_

\def\keyFont{\fontsize{8}{11}\helveticabold }
\def\firstAuthorLast{Hathaway {et~al.}} %use et al only if is more than 1 author
\def\Authors{David H. Hathaway\,$^{1,*}$, Lisa A. Upton\,$^{2}$ and Sushant S. Mahajan\,$^{1}$}
\def\Address{$^{1}$HEPL, Stanford University, Stanford CA 94305 USA \\
$^{2}$Southwest Research Institute, Boulder, CO, 80302 USA }
 \def\corrAuthor{D. H. Hathaway}

\def\corrEmail{dave.hathaway@comcast.net}

\begin{document}

\onecolumn
\firstpage{1}

\title[Axisymmetric Flow Variations]{Variations in Differential Rotation and Meridional Flow within the Sun`s Surface Shear Layer 1996-2022}

\author[\firstAuthorLast ]{\Authors} %This field will be automatically populated
\address{} %This field will be automatically populated
\correspondance{} %This field will be automatically populated

\extraAuth{}

\maketitle

\begin{abstract}

\section{}

We measure differential rotation and meridional flow in the Sun`s surface shear layer by tracking the motions of the magnetic network seen in magnetograms from SOHO/MDI and SDO/HMI over solar cycles 23, 24, and the start of 25 (1996-2022). We examine the axisymmetric flows derived from 15-24 daily measurements averaged over individual 27-day Carrington rotations.
Variations in the differential rotation include the equatorial torsional oscillation - cyclonic flows centered on the active latitudes with slower flows on the poleward sides of the active latitudes and faster flows equatorward. The fast flow band starts at $\sim$45$^\circ$ latitude during the declining phase of the previous cycle and drifts equatorward, terminating at the equator at about the time of cycle minimum. Variations in the differential rotation also include a polar oscillation above 45$^\circ$ with faster rotation at cycle maxima and slower rotation at cycle minima. The equatorial variations were stronger in cycle 24 than in cycle 23 but the polar variations were weaker.
Variations in the meridional flow include a slowing of the poleward flow in the active latitudes during cycle rise and maximum and a speeding up of the poleward flow during cycle decline and minimum. The slowing in the active latitudes was less pronounced in cycle 24 than in cycle 23. 
Polar countercells (equatorward flow) extend from the poles down to $\sim$60$^\circ$ latitude from time to time (1996-2000 and 2016-2022 in the south and 2001-2011 and 2017-2022 in the north).
Both axisymmetric flows vary in strength with depth. The rotation rate increases inward while the meridional flow weakens inward.

\tiny
 \keyFont{ \section{Keywords:} Solar Velocity Fields, Solar Photosphere, Solar Convection Zone, Rotation, Meridional Flow, Solar Cycle} %All article types: you may provide up to 8 keywords; at least 5 are mandatory.

 \end{abstract}

\section{Introduction}
\label{S-Intro} 
%%%%%%%%%%%%%%%%%%%%%%%%%%%%%%%%%%%%%%%%%%%%%%%%%%%%%%%%%%
The Sun's axisymmetric flows (differential rotation and the meridional circulation) are key aspects of the convection zone dynamics and are the principal drivers of the solar activity cycle.
The shearing motions of the differential rotation stretch north-south and radially oriented magnetic field into the azimuthal direction thereby producing strong toroidal field that eventually is buoyed up to produce the bipolar active regions of the solar activity cycle.
The meridional flow at the top of the surface shear layer transports magnetic field to the poles to reverse the polar fields at cycle maximum and build up new polar fields by cycle minimum.
Those polar fields appear to determine the strength of the next solar activity cycle \citep{Babcock61, Leighton69, Schatten_etal78, Svalgaard_etal05}.
The equatorward meridional flow deeper within the Sun may play a part in the equatorward drift of the active latitudes \citep{Choudhuri_etal95}.

These axisymmetric flows are likely produced by the effects of the Sun's rotation on the convective flows and thermal structure in the Sun's convection zone \citep{Busse70, BelvederePaterno76, GlatzmaierGilman82, Hotta_etal15, FeatherstoneMiesch15}.
However, \cite{Karak_etal15}, \cite{2019PhDT.......170M} and \cite{Hotta_etal22} have recently shown that Maxwell stresses associated with the Sun`s magnetic fields may also play an important part.
Variations in these flows occur as consequences of the interactions between the flows and the magnetic field and conversely the variations in the flow can have consequences for the magnetic field configuration itself.

Ideally we would like to know the variations in both differential rotation and meridional flow at all latitudes and depths within the convection zone over the course of many solar cycles.
This would help to answer key questions concerning how the variations result from solar activity (the effects of magnetic field on the flows) and how the variations influence solar activity (effects of the flows on the magnetic field).
In actuality we have measurements at a good range of latitudes and depths for just the last two cycles.
The differential rotation is fairly well characterized throughout the convection zone by global helioseismology (with the exception of a cone of uncertainty extending from the polar regions and expanding inward) \citep{Howe09}.
The meridional flow is an order of magnitude weaker and is therefore more poorly characterized - even at the photosphere - but more so deeper into the convection zone where the equatorward return flow must be \citep{Hathaway12C, Zhao_etal13}.

Here we examine variations in these flows over the course of two and a half solar cycles - cycles 23, 24, and the start of 25 - by measuring the flows at a small range of depths in the surface shear layer several times daily using magnetic pattern tracking on magnetograms from the Michelson Doppler Interferometer (MDI) instrument on the ESA/NASA Solar and Heliospheric Observatory (SOHO) spacecraft \citep{Scherrer_etal95} and from the Helioseismic and Magnetic Imager (HMI) instrument on the NASA Solar Dynamics Observatory (SDO) spacecraft \citep{Scherrer_etal12}.
The measurements start in May of 1996 and continue through May of 2022 with a short break in the summer of 1998 when radio contact with SOHO was temporarily lost.

\section{Measurement Method}
\label{S-Obs}
%%%%%%%%%%%%%%%%%%%%%%%%%%%%%%%%%%%%%%%%%%%%%%%%%%%%%%%%%%
The axisymmetric flows can be measured either by direct Doppler measurements of the photospheric plasma or by tracking photospheric features.
Tracked features include white-light intensity features (sunspots, faculae, and granules), spectroscopically derived features (H$\alpha$ filaments and the Ca II network), features identified by spectral line Doppler shifts (supergranules and the acoustic waves used in helioseismology), and features identified by spectral line Zeeman splitting (the magnetic network and its elements).

Each method and/or feature gives results for a given range of depths within the Sun.
Each has advantages and disadvantages depending on spatial coverage, accuracy, sources of systematic errors, and the flow being measured (differential rotation or meridional flow).
Sunspots were used as the earliest tracers of solar rotation and led to the discovery of the latitudinal differential rotation \citep{Carrington59B}.
Direct Doppler measurements of solar rotation produced the first indications of possible variations in the rotation rate \citep{Howard76} and led to the discovery of the torsional oscillations \citep{HowardLabonte80} - changes in rotation rate at given latitudes that vary with the 11-year period of the solar cycle.
 \cite{GilmanHoward84} used six decades (1921-1982) of sunspot measurements from the Mt. Wilson Observatory to find cycle related variations in rotation - an increase in rotation rate just after cycle maximum and another increase near minimum - which  are now recognized as manifestations of the torsional oscillation signal at high and low latitudes respectively.
Helioseismic measurements of the rotation profile with latitude and depth using global oscillation modes now show that the torsional oscillation signal extends deep into the convection zone \citep{Howe09}.

Measurements of the meridional flow have been much more difficult with even the direction of the flow being in question until the 1980s.
\cite{HowardLabonte81} found a poleward meridional flow of about 10 m s$^{-1}$ for photospheric magnetic features, with evidence of solar cycle related variations (slower early in the cycle and faster later).
\cite{Komm_etal93b} measured the movement of small photospheric magnetic features to find a similar meridional flow and further evidence of variations (slower at cycle maximum and faster at minimum).
In previous papers \citep{HathawayRightmire10, HathawayRightmire11, HathawayUpton14, Mahajan_etal21} we have adopted a similar method.

Here we measure the differential rotation and meridional flow by tracking the movement of the Sun's magnetic network with our MagTrak program.
Tracking the magnetic network pattern has several advantages.
These features cover the entire surface of the Sun and persist for hours to days.
Tracking their motions provide measurements of both the differential rotation and meridional flow at a wide range of latitudes and a range of depths within the Sun`s surface shear layer.
We do this by first mapping full disk magnetograms from SOHO/MDI and SDO/HMI onto heliographic longitude and latitude using a bi-cubic interpolator and then cross-correlating long ($\sim$105$^\circ$ in longitude) thin ($\sim$ 2$^\circ$ in latitude - the typical width of the supergranules that form the magnetic network) strips of the mapped data with strips from maps at later and earlier times separated by time lags ranging from 96 minutes to 8 hours.
We find the shift in longitude and latitude that maximizes the correlation between the two strips with a precision of about 0.1 pixel or less.

In mapping the full disk magnetograms to grids with points equispaced in both latitude and longitude we make several corrections to the image geometry based on studies of the one year overlap between SOHO/MDI and SDO/HMI as well as other reported image issues.
The location of disk center in MDI data was found to be different depending upon the orientation of SOHO with $(x_0,y_0) \rightarrow (x_0+0.45,y_0+0.90)$ when north is up and $(x_0,y_0) \rightarrow (x_0+2.10,y_0+1.00)$ when north is down.
(Note: we take the origin of the image to be the lower left corner of the lower left pixel rather than the center of that pixel.)
We correct the given latitude at disk center, $B_0$, and the position angle of the rotation axis, $P_0$, for the corrections to the orientation of the Sun's rotation axis given by \cite{BeckGiles05} along with $P_0 \rightarrow P_0-0.21$ prior to 2003 and $P_0 \rightarrow P_0-0.27$ after 2003.
We then correct the plate scale by taking the radius of the image $r_0 \rightarrow r_0\times0.9996$.
For HMI data we also take the origin to be at the lower left corner of the lower left pixel and correct $B_0$ and $P_0$ for the changes in the orientation of the Sun's rotation axis given by \cite{BeckGiles05}

Some studies have questioned the accuracy of our measurement technique.
\cite{Dikpati_etal10A} and \cite{Guerrero_etal12} examined the effects of diffusion on measurements of the movement of the magnetic pattern by producing artificial data in the form of a diffuse magnetic field with simple dipoles representing active regions, transporting those fields with diffusion and axisymmetric advection, and then measuring the motions.
They both conclude that there are large systematic errors in the magnetic feature tracking technique due to the effects of diffusion.
While neither of these studies describe precisely how the motions were measured (no mention of time lags or correlation windows) but they do suggest that we should see apparent motion away from the active latitudes.
We \citep{HathawayRightmire11}, tested our MagTrak program using artificial data that matched the characteristics of the magnetic network itself by advecting magnetic elements with an evolving supergranule flow field.
The resulting magnetic pattern is not diffuse but instead faithfully mimics the network pattern seen on the Sun.
Our measurements found no evidence of any apparent flows away from active latitude.
Furthermore, we see no evidence of flow away from the active latitudes in actual data as suggested by the effects of diffusion.
Recently however, a true source of systematic error has been discovered which was unanticipated by any of these previous tests. 
In the following section we describe how this error is removed in MagTrak 3.0 along with a number of other improvements to the program.

\section{Improved Measurements from Magnetic Network Feature Tracking}
\label{S-Improvements}
\begin{figure}[htb]
% \centering
\includegraphics[width=1.0\columnwidth]{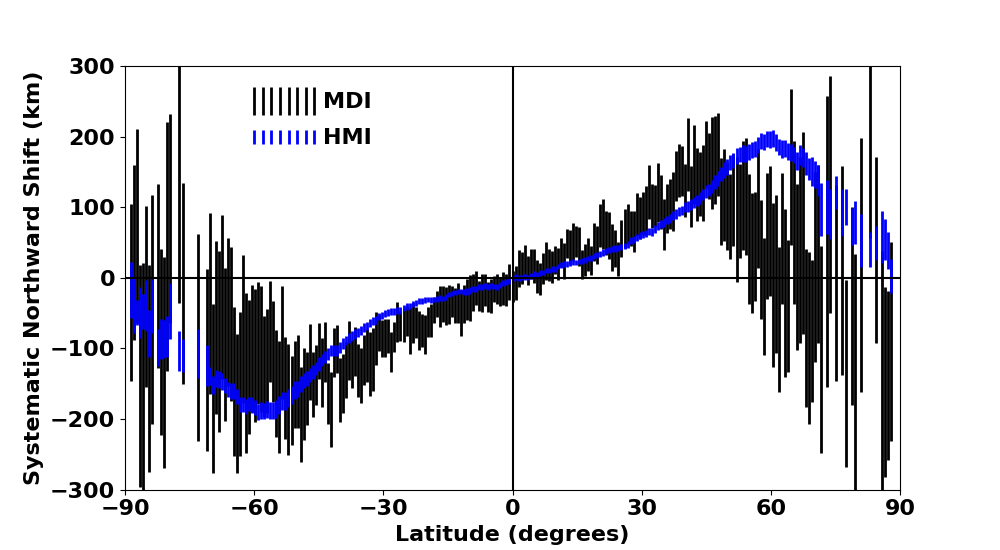}
\caption{The average (spurious) poleward shift as a function of latitude is shown in blue with 2$\sigma$ error bars for the HMI data and in black  with 2$\sigma$ error bars for the MDI era. This apparent shift away from disk center must be removed from meridional flow measurements.}
\label{fig:Shift}
\end{figure}

%%%%%%%%%%%%%%%%%%%%%%%%%%%%%%%%%%%%%%%%%%%%%%%%%%%%%%%%%%
Recently, \cite{Mahajan_etal21} found a spurious signal in flow measurements associated with tracking the Sun's magnetic network.
This spurious signal manifests itself as an apparent shift of the magnetic features away from disk center much like that seen for acoustic waves in time-distance helioseismology studies \citep{Zhao_etal12}
It has the interesting property of reaching a fixed shift within about an hour of solar rotation.
The magnitude of this spurious shift as a function of heliocentric angle from disk center can be determined by measuring the apparent shift using three different time lags as discussed in \cite{Mahajan_etal21}.
The correlation tracking measurements can then be corrected by removing this spurious shift from the measured shifts before dividing by the time lag to get flow velocities in m s$^{-1}$.
The spurious shift in the north-south direction determined using the HMI data and time lags of 2, 4, and 8 hours is shown in blue in Fig. \ref{fig:Shift} while the shift determined using MDI data with time lags of 96, 192, and 480 minutes is shown in black.
Note that a shift of 150 km gives a spurious poleward velocity of 5 m s$^{-1}$ for an 8 hour time lag but increases to 20 m s$^{-1}$ for a 2 hour time lag.
This spurious shift has influenced previous measurements, introducing large errors for time lags less than 8 hr and small but significant errors for longer time lags.
The source of this signal is still somewhat uncertain but most likely related to the spectral line depth of formation at different angles from disk center.
The slight differences between the profiles for MDI and HMI might be attributed to the different spectral lines used along with the differences in spatial and spectral resolution.

Further improvements on the measurement method as described in \cite{Mahajan_etal21} include broadening the displacement search area for finding the maximum in the cross correlation.
This is to ensure that we are sampling the full range of displacements.
We also iteratively shift the data strips by fractional pixels to avoid the ``peak locking'' problem associated with the tendency of these methods to avoid giving half pixel shifts.

We have measured the differential rotation and meridional flow profiles at 255 equi-spaced latitude positions from pole to pole at 96 minute intervals with the MDI instrument and at 60 minute intervals with the HMI instrument.
The several hundred individual measurements made over each 27-day Carrington rotation are averaged and standard errors are calculated.
While the 7$^\circ$ tilt of the Sun`s rotation axis relative to the plane of the Earth`s orbit allows us to see all the way to its north pole in September of each year and all the way to its south pole in March of each year, the foreshortening of our view near the limb makes the polar measurements most uncertain.

\begin{figure}[htb]
% \centering
\includegraphics[width=1.0\columnwidth]{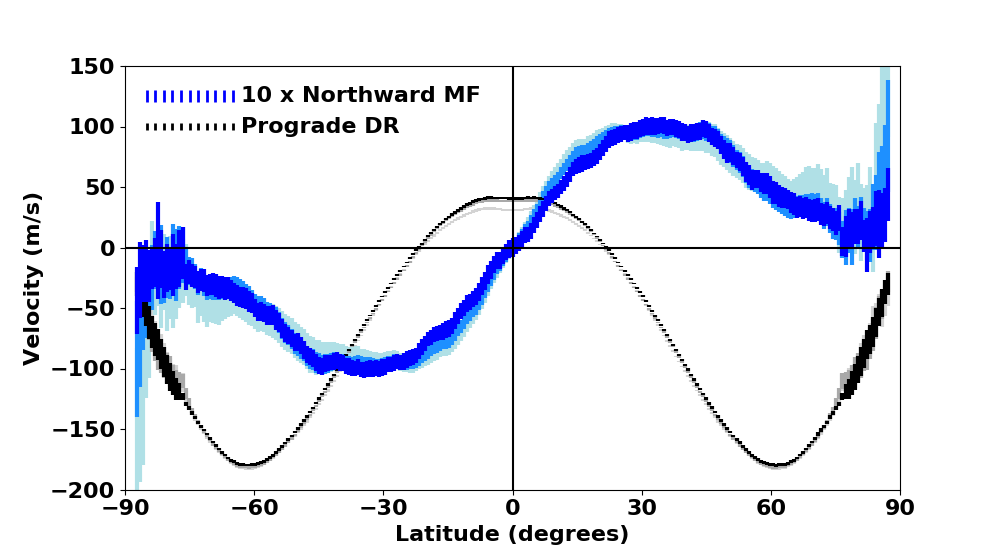}
\caption{The average axisymmetric flows. The north/south symmetric differential rotation (positive in the prograde direction) is shown in black with 2$\sigma$ error bars for 8-hour time lags, in gray for 4-hour time lags, and in light gray for 2-hour timelags. The north/south anti-symmetric meridional flow (multiplied by 10, positive in the northward direction) is shown in dark blue with 2$\sigma$ error bars for 8-hour time lags, in medium blue for 4-hour time lags, and in light blue for 2-hour time lags.}
\label{fig:AveFlows}
\end{figure}

The differential rotation and meridional flow profiles averaged over nearly 100,000 individual profiles measured with 8-hour, 4-hour, and 2-hour time lags over the 11 years from May 2010 through April 2021 are shown in Fig. \ref{fig:AveFlows}.
The differential rotation profile has a flattened peak at the equator with a slight dip right at the equator.
The latitudinal velocity shear reaches a maximum at about 30$^\circ$.
The meridional flow profile reaches its maxima of 9 m s$^{-1}$ at about 30$^\circ$.
The high latitude meridional flow now appears to drop to zero at about 80$^\circ$ instead of going all the way to the poles.
This aspect of the meridional flow profile only becomes apparent with the removal of the spurious shift away from disk center.

The equatorial differential rotation velocity increases with longer time lags - 30.7 m s$^{-1}$ at 2-hours, 38.1 m s$^{-1}$ at 4-hour time lags, and 40.4 m s$^{-1}$ at 8-hour time lags.
The meridional flow speed decreases with increasing time lag but not as much as was reported in \cite{Hathaway12C}.
This is directly attributed to the effects of the spurious systematic shift on the short time lag velocities.
The 8-hour time lag average profiles are removed from the individual Carrington rotation profiles to reveal the variations in these flows over solar cycles 23, 24, and the start of 25.

\section{Variations in the Differential Rotation}
\label{S-DR}
%%%%%%%%%%%%%%%%%%%%%%%%%%%%%%%%%%%%%%%%%%%%%%%%%%%%%%%%%%

\begin{figure}[htb]
% \centering
\includegraphics[width=1.0\columnwidth]{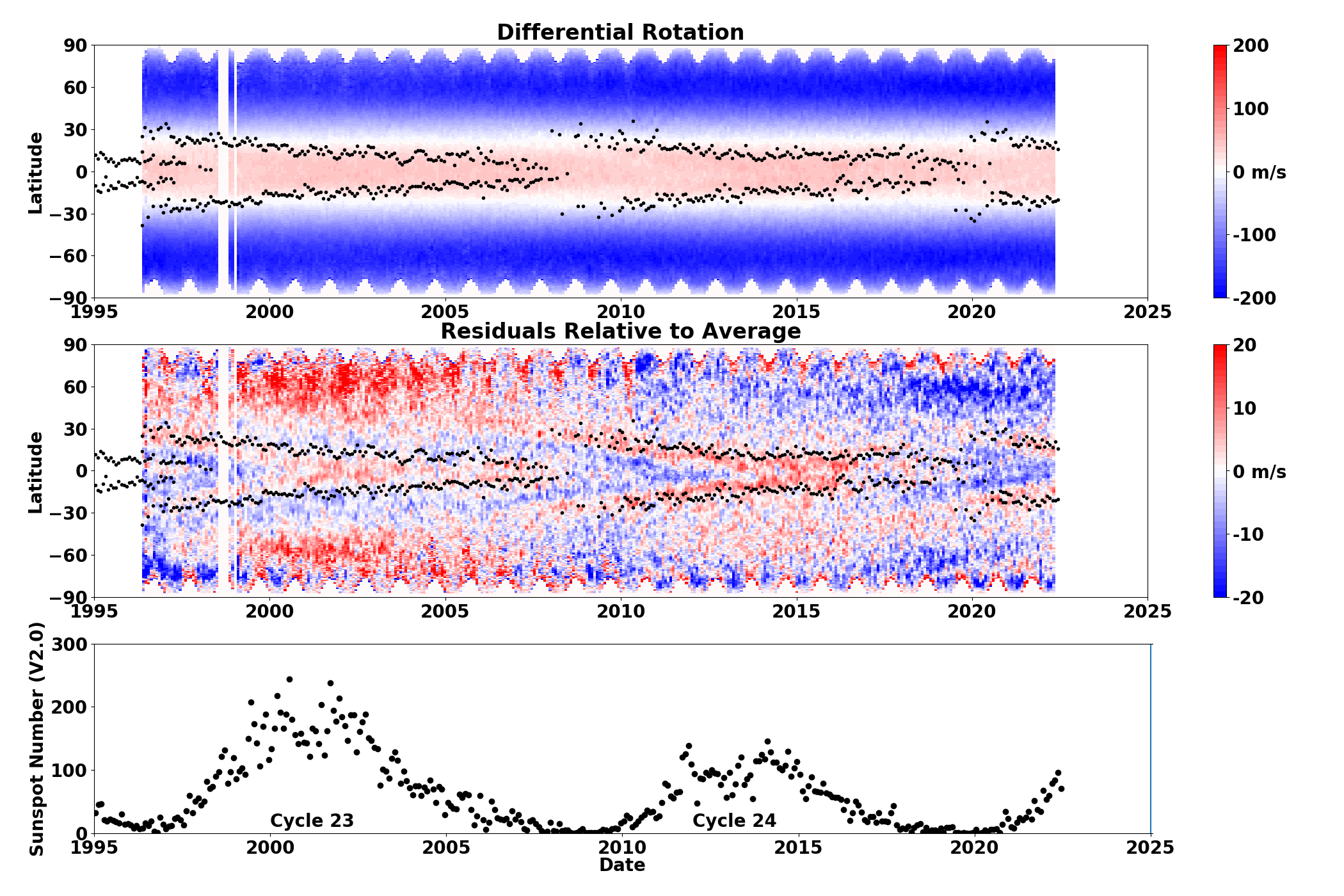}
\caption{Differential rotation (east-west flow relative to the frame of reference rotating at the Carrington rate and positive in the direction of rotation) as a function of time and latitude is shown in the upper panel.
The residual east-west flow found by removing the north/south symmetric average profile is shown in the central panel.
The latitudes of the centroids of sunspot areas for each Carrington rotation are shown by the black dots in each upper panel.
The monthly average of the daily sunspot number (V2.0) is shown for reference in the bottom panel.}
\label{fig:DRhistory}
\end{figure}

The history of the Carrington rotation averaged profiles of the differential rotation is shown in the top panel of Fig. \ref{fig:DRhistory}.
The history of the variations from the average profile is shown in the central panel.
The monthly sunspot numbers (V2.0) are shown for reference in the bottom panel to illustrate the progression of the sunspot cycle.
The small black dots in each of the two upper panels mark the locations of the sunspot area centroids in each hemisphere for reference to the locations of the active latitudes.

Variations in the differential rotation are at a level less than about 5\% of the flow speed itself.
This is evident in the lack of any obvious variations in the full profile history shown in the top panel of Fig.  \ref{fig:DRhistory}.
The variations seen in the residuals relative to the average shown in the central panel of Fig. \ref{fig:DRhistory} have two primary components, one at low latitudes and one at high latitudes.
These components may, or may not, be connected.
The torsional oscillation (an 11-year oscillation of faster and then slower flow in the active latitudes, first reported by \cite{HowardLabonte80}) is seen straddling the sunspot zones in the center panel of Fig.  \ref{fig:DRhistory}.
Faster than average flows are seen on the equatorward side of the active latitudes delineated by the black dots at the latitudes of the sunspots area centroids in each hemisphere.
These faster than average flows can be traced to higher latitudes and earlier times, giving rise to the recognition of an extended cycle with several years of overlap between adjacent cycles \citep{MartinHarvey79, Snodgrass87A, Wilson_etal88}.
The slower than average flows are seen on the poleward sides of the sunspot zones and become strongest near the equator after the sunspots have disappeared.
The second component presents as an oscillation at latitudes above about 45$^\circ$ with faster rotation at times of cycle maxima and slower rotation at times of minima.

Both of These oscillating components vary significantly from cycle to cycle.
The faster rotation near the equator was stronger in cycle 24 than it was in cycle 23.
The faster rotation at high latitudes near cycle maximum was much stronger in cycle 23 than it was in cycle 24 while the slower rotation at high latitudes near cycle minimum was much stronger at cycle 24/25 minimum than it was at cycle 23/24 minimum.

We also see significant north/south differences in these flow variations.
The faster rotation at high latitudes near cycle maximum was stronger in the north in cycle 23 but stronger in the south in cycle 24.
Notably, this increase in rotation in the north for cycle 24, while present, is almost imperceptible in Fig. \ref{fig:DRhistory}.
The slower rotation at high latitudes near cycle minimum was stronger in the south at cycle 23/24 minimum but stronger in the north at cycle 24/25 minimum.
Similar cycle-to-cycle and hemispheric differences in the differential rotation as well as other features seen in Fig. \ref{fig:DRhistory} have also been reported by \cite{Lekshmi_etal18} and \cite{Getling_etal21} from helioseismic studies.

\begin{figure}[htb]
 \centering
\includegraphics[width=1.0\columnwidth]{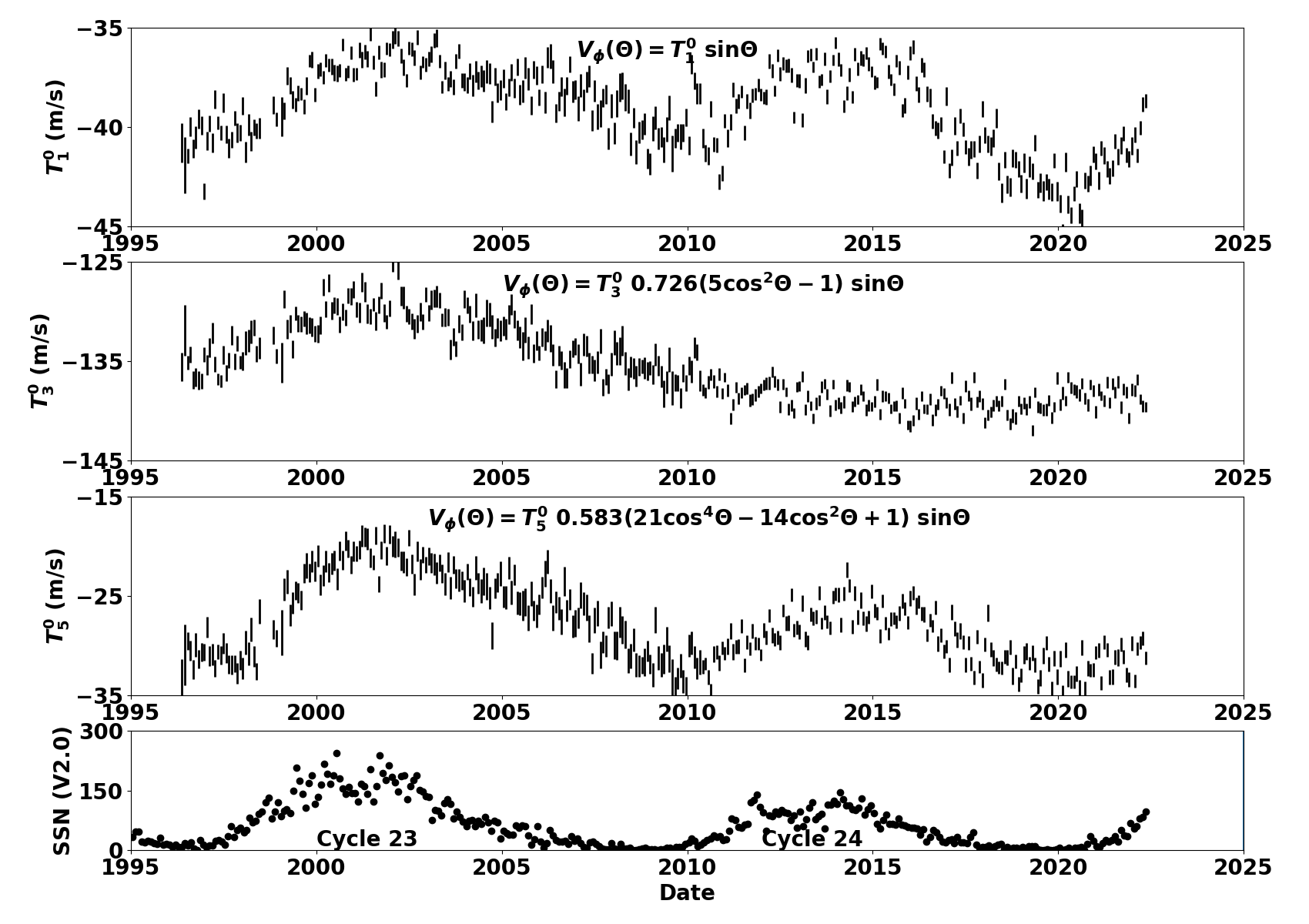}
\caption{Legendre polynomial fit coefficient histories for the differential rotation.
The top panel shows the $T^0_1$ coefficient which gives the solid body rotation relative to the Carrington rotation frame of reference.
Values for each Carrington rotation are shown with 2$\sigma$ error bars.
The second and third panels show the $T^0_3$ and $T^0_5$ coefficients respectively.
The monthly average of the daily sunspot number (V2.0) is shown for reference in the bottom panel.}
\label{fig:DRhistoryLegendre}
\end{figure}

The variations in the differential rotation can be quantified by fitting each latitudinal profile with vector spherical harmonics.
The coefficient histories for the first three axisymmetric and north/south symmetric components are shown in Fig.  \ref{fig:DRhistoryLegendre} along with the monthly sunspot numbers for reference.

The $T_1^0$ coefficients shown in the top panel give the solid body rotation relative to the Carrington rotation frame of reference.
They exhibit clear solar cycle related variations with faster rotation at cycle maxima and slower at cycle minima.
These variations clearly must represent a periodic radial transfer of angular momentum between different layers within the Sun`s convection zone.
The rotation rate peak in cycle 24 was only slightly smaller than that for the much larger cycle 23.
The rotation rate minimum at cycle 24/25 minimum was significantly slower than at the previous two minima.

The $T_3^0$ and $T_5^0$ coefficients characterize the differential rotation.
While $T_5^0$ appears to be directly related to the solar activity cycle, $T_3^0$ indicates a much stronger (more negative) latitudinal shear in cycle 24.
This difference can be attributed to the weaker polar spin-up seen in cycle 24 in Fig. \ref{fig:DRhistory}.

The $T_2^0$ coefficient (not shown here) characterizes the north/south asymmetry in the differential rotation.
It shows a faster northern hemisphere throughout cycle 23 and a faster southern hemisphere throughout cycle 24 and into cycle 25.

These variations in differential rotation are thought to be consequences of feedback from the magnetic structures of the activity cycle (active latitudes and polar fields).
This includes flows associated with thermal structures \cite{Spruit03} and Maxwell stresses associated with the magnetic field itself \cite{Schussler81, Yoshimura81, Rempel12}.
The variations we observe may help to further constrain the mechanisms involved.

\section{Variations in the Meridional Circulation}
\label{S-MFx}
\begin{figure}[htb]
\centering
\includegraphics[width=1.0\columnwidth]{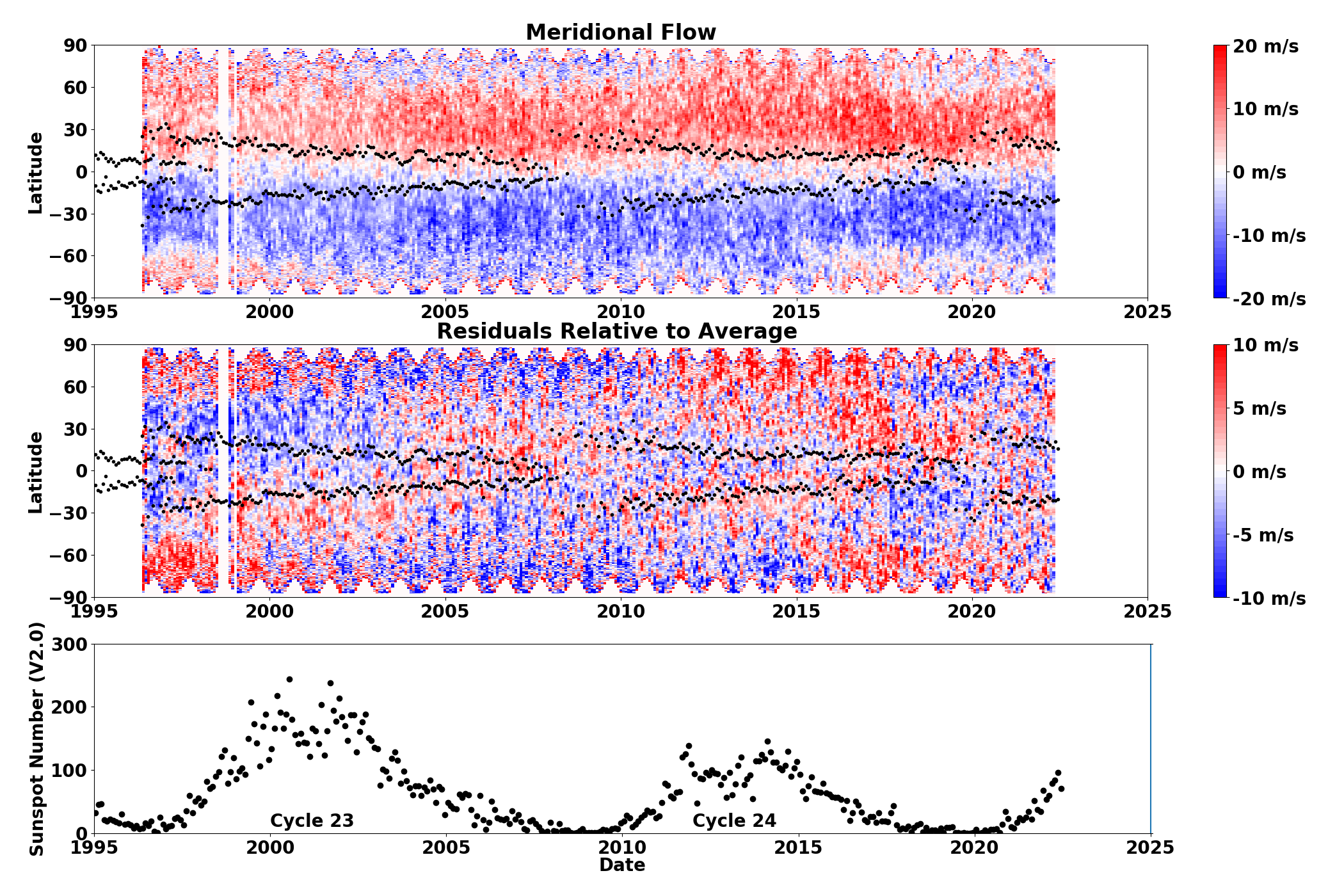}
\caption{Meridional flow (north-south flow, positive northward) as a function of time and latitude is shown in the upper panel.
The residual meridional flow found by removing the north/south antisymmetric average profile is shown in the central panel.
The latitudes of the centroids of sunspot areas for each Carrington rotation are shown by the black dots in each upper panel.
The monthly average of the daily sunspot number (V2.0) is shown for reference in the bottom panel.}
\label{fig:MFhistory}
\end{figure}

%%%%%%%%%%%%%%%%%%%%%%%%%%%%%%%%%%%%%%%%%%%%%%%%%%%%%%%%%%
The history of the Carrington rotation averaged profiles of the meridional flow is shown in the top panel of Fig. \ref{fig:MFhistory}.
The history of the residual variations relative to the average profile is shown in the central panel.
The monthly sunspot numbers (V2.0) are shown in the bottom panel to illustrate the progression of the sunspot cycle for reference.
The small black dots in each of the two upper panels mark the locations of the sunspot area centroids in each hemisphere for reference to the locations of the active latitudes.

The variations in the meridional flow are relatively large and are apparent even in the history of the individual flow profiles (top panel in Fig. \ref{fig:MFhistory}).
The residuals relative to the average profile (central panel in Fig. \ref{fig:MFhistory}) show a slowing of the poleward meridional flow in both hemispheres during the rise and maximum of cycle 23.
This slowing during rise and maximum is much less evident in cycle 24 and 25 and appears to be concentrated in the sunspot zones.
Note that this is a slowing of the meridional flow at a range of latitudes, including latitudes on the equatorward sides of the sunspot zones.
We do not see inflows toward the sunspot zones.
Inflows would be characterized by an increased poleward flow on the equatorward sides of the sunspot zones which is not present here.
(This is likely a consequence of the depth associated with these measurements - the middle of the surface shear layer rather than nearer to the surface.)
The poleward flow increases during the declining phase of each cycle and the approach to minimum.
This increase was stronger in cycle 24 than it was in cycle 23.

In addition to the changes in the poleward flow seen in the lower, active latitudes, we also see polar countercells - equatorward flow from the poles down to about 60$^\circ$ latitude appearing from time to time.
These countercells can be seen in the full profiles shown in the upper panel of Fig. \ref{fig:MFhistory}.
We see a counter cell in the south fully established at the start of the dataset in 1996.
This counter cell shrinks and disappears by 2001/2002.
As this southern countercell disappears, a countercell forms in the north, extends down to 60$^\circ$ by 2006 and then shrinks and disappears by 2011.
From 2011 to 2015 there are no countercells but a cell forms in the south in 2016 and then in the north in 2017 and both continue to exist until the end of the dataset in mid-2022.
These countercells are likely real features.
They persist for years but come and go without any apparent connection to instrumental changes, spacecraft orientation, or solar magnetic fields.
The presence of these countercells and the low latitude variations in the meridional flow can have consequences for the solar dynamo \citep{Dikpati_etal04, Karak10}.

\begin{figure}[htb]
\centering
\includegraphics[width=1.0\columnwidth]{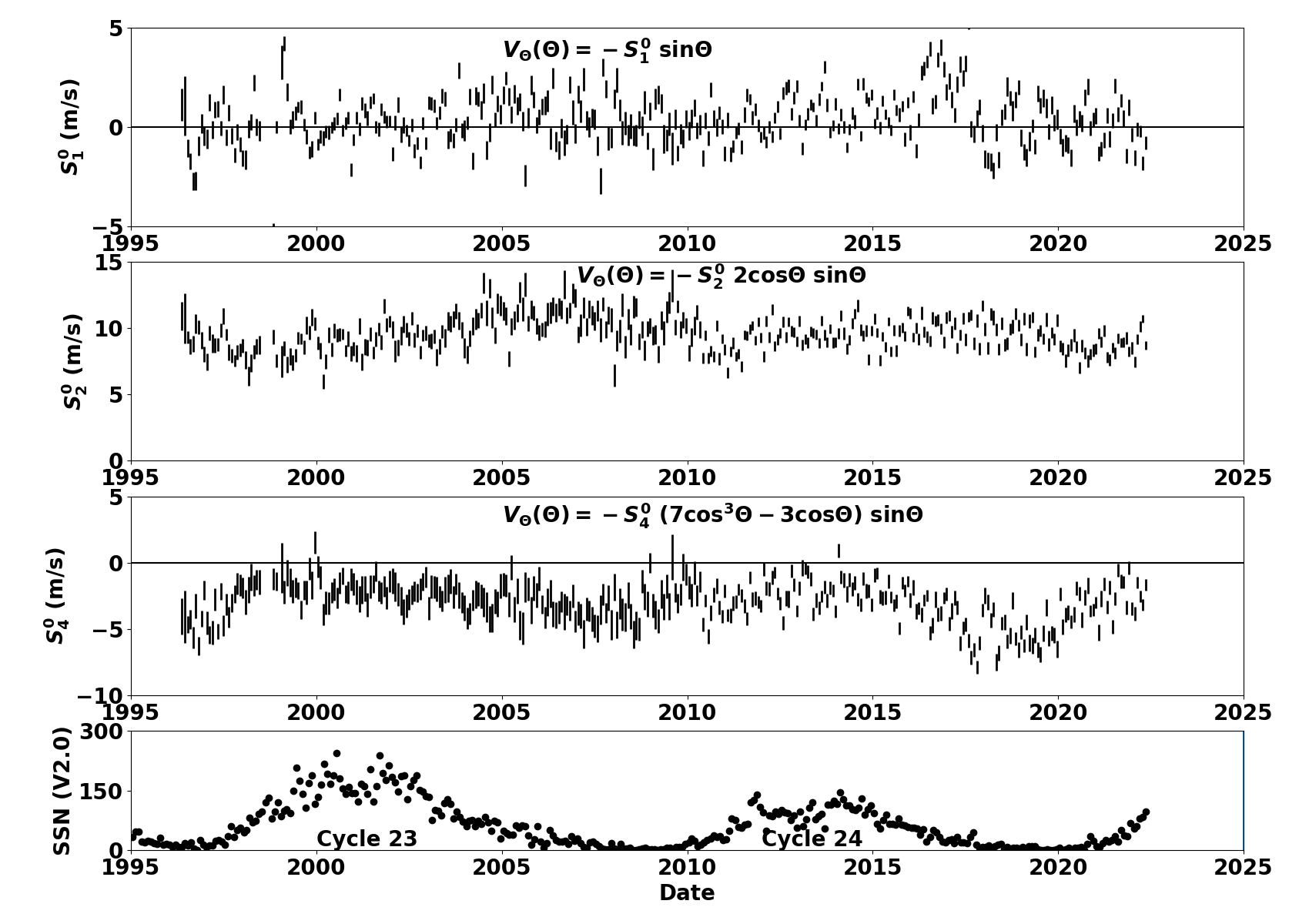}
\caption{Legendre polynomial fit coefficient histories for the meridional circulation.
Values for each Carrington rotation are shown with 2$\sigma$ error bars.
The top panel shows the $S^0_1$ coefficient which gives a meridional flow with one cell from pole to pole (positive toward the north).
The second panel shows the $S^0_2$ coefficient which gives two cells from pole to pole with positive values indicating poleward flow in each hemisphere.
The third panel shows the $S^0_4$ coefficient which gives four cells from pole to pole with negative values giving flow away from the poles at high latitudes.
The monthly averages of the daily sunspot number (V2.0) are shown for reference in the bottom panel.}
\label{fig:MFhistoryLegendre}
\end{figure}

The variations in the meridional flow can also be quantified by fitting each latitudinal profile with vector spherical harmonics.
The coefficient histories for three key components are shown in Fig.  \ref{fig:MFhistoryLegendre} along with the monthly sunspot numbers for reference.

The $S_1^0$ coefficients in the top panel represent a single meridional circulation cell extending from pole to pole with positive values indicating flow to the north across the equator.
This coefficient is near zero on average (0.4 m s$^{-1}$) but does indicated small but significant variations that give both cross-equator flow and north/south asymmetry to the poleward meridional flow.
Positive values indicate stronger poleward flow in the northern hemisphere.

The $S_2^0$ coefficients in the second panel characterize the dominant two cell meridional circulation with positive values indicating poleward flow in each hemisphere.
The average amplitude of this flow component is 9.6 m s$^{-1}$.
These coefficients show the slowing of the poleward flow during the rising phase and maximum of cycle 23 with a less pronounced slowing during the rise of cycle 24.

The $S_4^0$ coefficients in the third panel characterize the four cell meridional circulation with positive values indicating poleward flow at high latitudes and equatorward flow at low latitudes.
The average amplitude of -3.0 m s$^{-1}$ is enough to effectively stop the poleward meridional flow at about 80$^\circ$ latitude and to push the peak in the average meridional flow profile from 45$^\circ$ down to about 30$^\circ$ as shown in Fig. \ref{fig:AveFlows}.
The more negative values starting in 2017 are large enough to produce the countercells in both hemispheres.

\begin{figure}[htb]
\centering
\includegraphics[width=1.0\columnwidth]{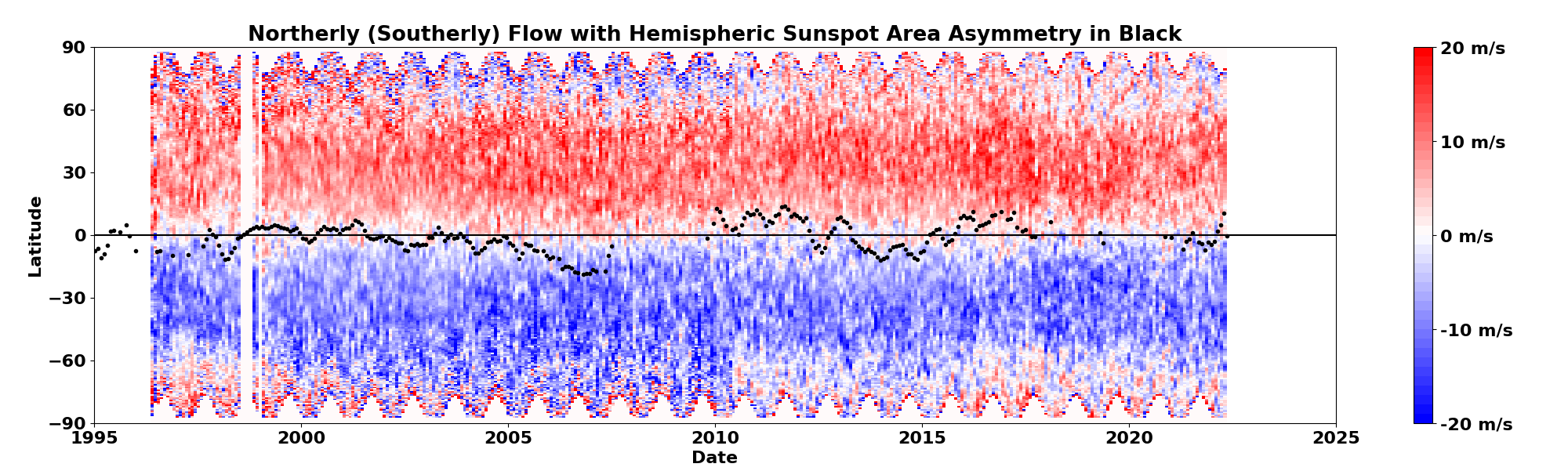}
\caption{The meridional flow and hemispheric asymmetries in sunspot area.
The asymmetry in hemispheric sunspot area, (North - South)/(North + South), is shown multiplied by 30 with the black dots near the equator.}
\label{fig:CrossEq}
\end{figure}

The cross equator flow holds some interest in terms of how it might be related to hemispheric differences in solar activity \citep{Komm22}.
In Fig \ref{fig:CrossEq} we superimpose a measure of hemispheric asymmetry - the difference in sunspot area between the two hemispheres divided by the sum of the sunspot areas.
We find periods (2012-2016) when there is flow across the equator away from the dominant hemisphere (northward flow away from the stronger south and southward flow away from the stronger north) but there are also times when the opposite is seen (2017).
A statistical test of the relationship between cross equator flow and hemispheric sunspot area asymmetry indicates a weak, but significant, anti-correlation - i.e. flow across the equator away from the dominant hemisphere.
This flow away from the more active hemisphere is opposite to the flows toward the active hemisphere found by \cite{Komm22}.
The likely explanation for this discrepency is the difference in flows around active regions at different depths - inflows near the surface and outflows at greater depths.

\section{Variations with Depth}
\label{S-Depth}
%%%%%%%%%%%%%%%%%%%%%%%%%%%%%%%%%%%%%%%%%%%%%%%%%%%%%%%%%%

As previously described, We make measurements of the axisymmetric flows using three different time lags for both the MDI and the HMI data.
This is necessary for our determination of the systematic shift but it also provides us with information about the flows at various depths within the surface shear layer.
Measurements at longer time lags are dominated by the magnetic pattern associated with larger, longer-lived convection cells that extend to greater depths within the Sun.
We can estimate depths associated with each time lag by comparing the equatorial rotation rate to that found from global helioseismology at depths within the surface shear layer.
This comparison (cf. \cite{Hathaway12C}) indicates that the measurements made with the 8-hour time lags represent flows about 25 Mm deep, while the 4-hour time lag measurements represent 22 Mm deep, and the 2-hour time lag measurements represent 18 Mm deep.

\begin{figure}[htb]
\centering
\includegraphics[width=1.0\columnwidth]{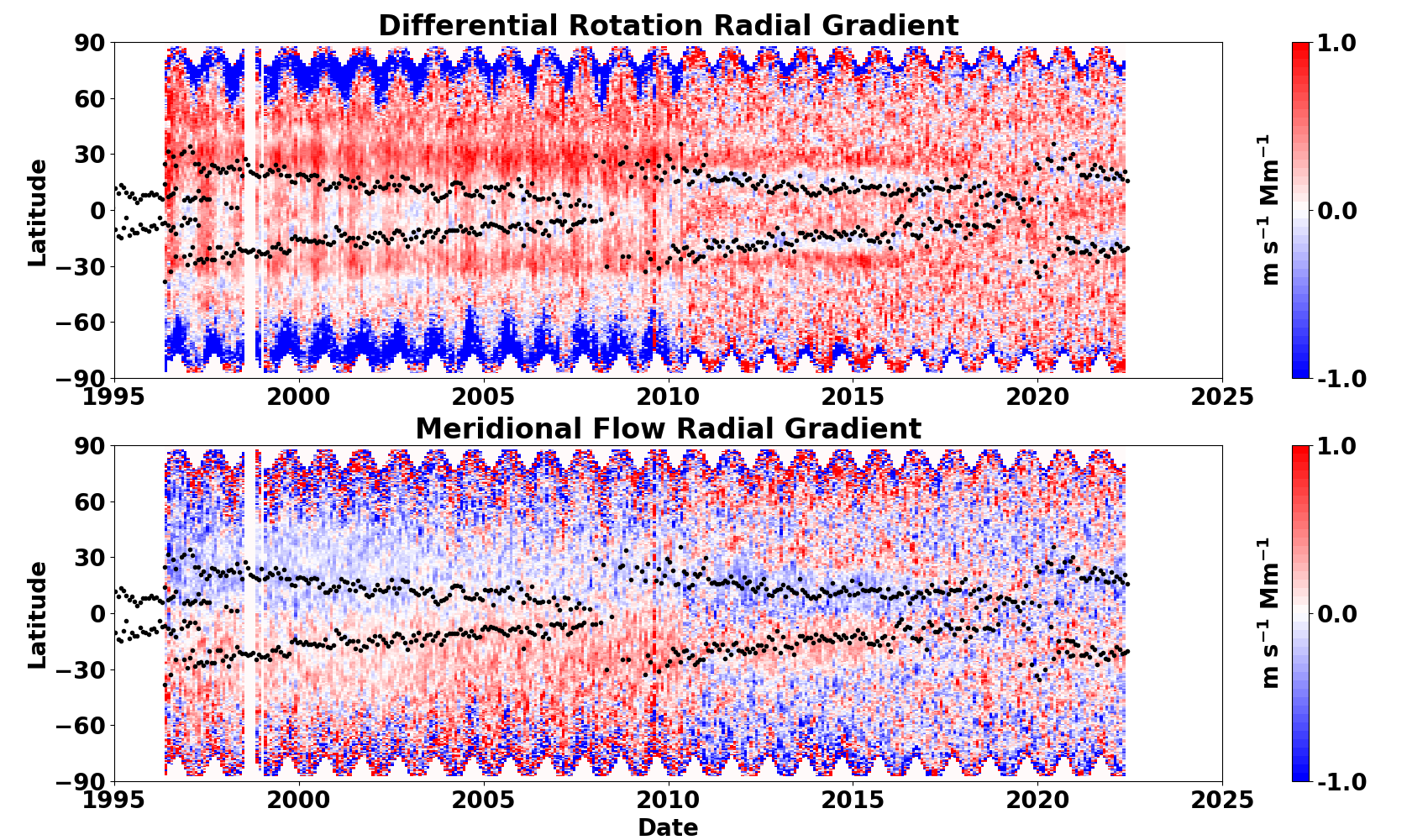}
\caption{The radial gradients in differential rotation (top panel) and meridional flow (bottom panel) in m s$^{-1}$ Mm$^{-1}$.
Positive values for the differential rotation indicate faster rotation going inward.
Positive values for the meridional flow indicates faster northward flow going inward.}
\label{fig:GradientHistory}
\end{figure}

Subtracting the differential rotation and meridional flow profiles at shorter time lags from those measured at 8-hour time lags and dividing by the difference in depth gives a measure of the radial gradients in these flows
Fig. \ref{fig:GradientHistory} shows these gradients using the 4-hour time lag measurements from HMI and the 192-minute time lag measurements from MDI as the shorter time lags.
Virtually the same results are found using the even shorter time lags - 2-hours for HMI and 96-minutes for MDI.

These radial gradient measurements show that the rotation rate increases inward while the meridional flow speed decreases inward.
Fig. \ref{fig:GradientHistory} shows that, in addition to these well known variations with depth, there are variations in these gradients with both time and latitude.
It also reveals artifacts which we don't fully understand.
The MDI era data suggests that the rotation rate gradient is stronger in the north than in the south while the HMI era data indicates that they are very similar.
The MDI era data shows an annual increase in the gradient at nearly all latitudes while this signal is absent from the HMI era data.
This is an obvious artifact.
Both MDI and HMI data show an enhanced gradient in the rotation rate at about 28$^\circ$.
This enhancement
does not drift equatorward with the active latitudes, but fades away during the HMI era - suggesting that it may not be an artifact.

The radial gradients in the differential rotation do not reveal the torsional oscillation signal.
This supports the findings from helioseismology indicating that these flow variations extend relatively unchanged through the surface shear layer \citep{Howe_etal00}.

The radial gradient in the meridional flow also has surprises.
During the MDI era the slowing of the meridional flow is evident at all latitudes.
During the HMI era the slowing of the meridional flow appears to be concentrated in the active latitude bands for both cycle 24 and 25. 

\section{Conclusions}
\label{S-Conclude}

\begin{figure}[htb]
\centering
\includegraphics[width=1.0\columnwidth]{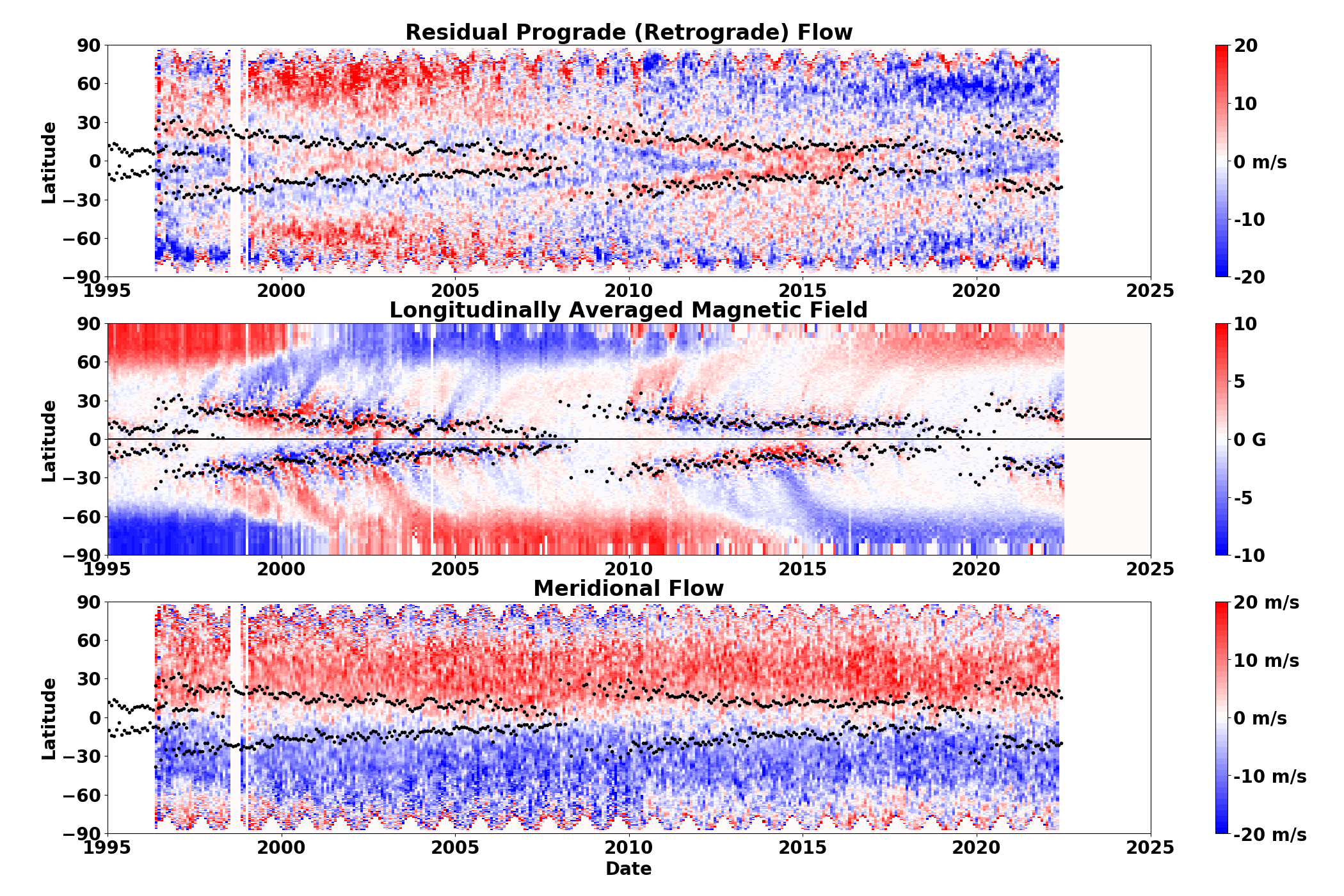}
\caption{Axisymmetric flow variations and associated magnetic field variations.
The differential rotation flow variations are shown in the top panel. The meridional flow is shown in the bottom panel.
The evolution of the Sun`s surface magnetic field averaged over longitude for each Carrington rotation is shown for reference in the central panel.}
\label{fig:MagBfly}
\end{figure}

We have used our improved MagTrak program for measuring the differential rotation and meridional flow indicated by the motions of the magnetic network features to determine the latitudinal profiles of the flows several times each day from mid-1996 through mid-2022.
This dataset covers the entirety of cycles 23 and 24 and catches the rise of cycle 25.
Our findings confirm several aspects of these flows and their variations and introduce new features.
Fig. \ref{fig:MagBfly} shows the axisymmetric flow variations along with the longitudinally averaged magnetic field to help guide our conclusions.

Our measurements of the differential rotation find the torsional oscillations associated with the sunspot zones and their extensions to higher latitudes earlier than the appearance of sunspots.
The cyclonic nature of these flow variations are consistent with either their association with inflows toward the sunspot zones \citep{Spruit03} or Maxwell stresses due to the presence of magnetic fields \citep{Schussler81, Yoshimura81, Rempel12} and would suggest that they are a response to the activity and not a cause of it.
However, we find that the active latitude variations were stronger in cycle 24 than in cycle 23 in spite of the fact that cycle 24 was the weaker cycle.
We might expect stronger torsional oscillation flows to be associated with the stronger inflows or magnetic structures in the more active cycle.
We also find a weak association with cross equator flow directed away from the more active hemisphere - an indication of outflow from, not inflow to, the activity bands at the depths we probe.
The apparent lack of any signal associated with the torsional oscillation in the radial gradient data shown in Fig. \ref{fig:GradientHistory} suggests the torsional oscillations flows don`t change with depth. Yet the change from inflows to outflows with depth - if that is the source - would suggest a more significant change.

The high latitude spin-up at cycle maxima is most prominent in the north in cycle 23.
It is seen weakly in the south in cycle 24 but is virtually nonexistent in the north in that cycle.
The spin-ups don`t appear to be associated with changes in the meridional flow - a faster poleward flow might tend to spin up the poles but we see a slowing of the high latitude meridional flow at the maxima in cycle 23 and a speed up at high latitudes in the north in cycle 24 - the opposite to what would be expected if the meridional flow was the source of this signal.
These observations suggest that neither the low latitude nor the high latitude torsional oscillations are caused by the Coriolis force acting on meridional flows.

The meridional flow is known to play a dominant role in transporting magnetic fields poleward as seen in a multitude of surface magnetic flux transport studies \citep{Jiang_etal14, Wang17}.
We now find that the meridional flow tends to disappear above about 80$^\circ$ - a feature employed ad hoc in some surface flux transport models. 
We also find that the speed of the meridional flow varies by 10-20\% and, more importantly, exhibits polar countercells from time to time.
The effects of these meridional flow variations are not immediately obvious when comparing the meridional flow and the magnetic field histories shown in Fig. \ref{fig:MagBfly}.
Further experimentation with these flows in surface flux transport models are needed to explore these connections.

\section{Acknowledgements}

DHH and SSM are supported by NASA contract NAS5-02139 (HMI) to Stanford University.
LAU was supported by NASA Heliophysics Living With a Star grants NNH16ZDA010N-LWS and NNH18ZDA001N-LWS and by NASA grant NNH18ZDA001N-DRIVE to the COFFIES DRIVE Center managed by Stanford University. 
HMI data used in this study are courtesy of NASA-SDO and the HMI science team. MDI data used in this study are courtesy of NASA/ESA-SOHO and the MDI science team.
%%%%%%%%%%%%%%%%%%%%%%%%%%%%%%%%%%%%%%%%%%%%%%%%%%%%%%%%%%

%%% BIBLIOGRAPHY %%%%%%%%%%%%%%%%%%%%%%%%%%%%%%%%%%%%%%%%%
%\section{References} 
\bibliographystyle{Frontiers-Harvard}
\bibliography{Hathaway}  

\end{document}